# Controlling solar radiation forces with graphene in plasmonic metasurface

Sina Soleymani[1,*], Sevda Seyyedmasoumian,[2] Asma Attariabad,[2] Sepideh Soleymani,[3] Farzaneh Bayat,[4] and Hamid Sabet[1,*]

[1] *Massachusetts General Hospital, Harvard Medical School, Boston, MA, 02129, USA*
[2] *Faculty of Electrical and Computer Engineering, University of Tabriz, Tabriz, Iran*
[3] *Faculty of New Sciences and Technologies, University of Tehran, Tehran, Iran*
[4] *Department of Physics, Azarbaijan Shahid Madani University (ASMU), Tabriz, 53751-71379, Iran*

**Email**:
Hsabet@mgh.harvard.edu, Ssoleymani1@mgh.harvard.edu

**Abstract**

Controlling and harvesting solar radiation pressure is a significant challenge, however, there are few potential solutions, which are suitable for several key applications. In this study, an electrically tunable plasmonic metasurface is designed for the visible spectrum. Moreover, the normal and the tangential optical forces acting on the metasurface are calculated. Whilst presenting high efficiency in the anomalous reflection, the designed active metasurface provides tunability of optical forces acting on the metasurface. The metasurface is composed of tapered silver cells embedded on top of the graphene layer with 20 layers of graphene sheets. Hence, by tuning the Fermi level of graphene sheets, the transferred momentum to the metasurface can be controlled. Our results can provide a suitable platform for optical force control desired in tunable radiation pressure harvesting, micro vehicles, solar sailing and optical tweezers.

## 1. Introduction

Metasurfaces with anomalous reflection/refraction are widely used for wavefront shaping by engineering the phase of light through optically thin phase gradient unit cells [1], [2]. The exotic anomalous response of light has found its applications in flat lenses, holography, and waveplate designs, recently [3], [4]. From many different phase gradient metasurface configurations, the phase gradients created by plasmonic nanoresonators are used for applications in visible and infrared light regions and are referred to as plasmonic metasurfaces [5]. The light-matter interactions and specifically momentum handover in these peculiar surfaces are fields of interest for many researchers, recently [6]. A metasurface due to anomalous reflection/refraction can produce anomalous momentum transfer from the light source to the surface [7].

Harvesting radiation pressure can become feasible by designing metasurfaces and wavefront engineering, which has applications in nanosailing, nanoparticle manipulation, and microscopic optical metavehicles [8-12], [37-38]. Considering the angle of incident and reflected light, the momentum transferred from photons to the surface can have normal and tangential components. Having a tangential component of the optical force is unfeasible on specular surfaces since the incident angle is equal to the reflected angle [7]. On the other hand, the time-averaged pressure acting on a perfectly absorbing surface by a normal incident can be written as $\langle P \rangle = I/c$, while this value becomes twice bigger for a perfectly reflecting specular surface $\langle P \rangle = 2I/c$, where $I$ and $c$ are intensity and speed of light, respectively, which demonstrates the significance of surface behavior in radiation pressure harvesting [13]. Therefore, by having metasurface configurations with tunable absorption/reflection one can control the exerted pressure magnitude. On the other hand, in anomalous reflectors the facility to actively modify the ratio of the anomalous reflection to specular reflection gives us the capability to control the amount of the tangential and normal forces. In this research we are going to exhibit that the alteration of force and its direction is possible by simply altering the anomalous reflection efficiency of the metasurface.

Since the tunability of the light-matter interactions is highly desired in photonics, embedding metasurfaces with alterable and active materials such as graphene, liquid crystals, and nonlinear materials such as Lithium Niobate are appealing subjects of research [14-18]. Graphene is an outstanding 2D material, with a unique band structure. The chemical potential level of a graphene sheet can be easily manipulated via gating or doping, which in turn changes its optical properties [19]. Thus, graphene is an enticing candidate for integration with metasurfaces, due to its ability to turn many optical configurations into an active device with applications in light-sailing, beam deflecting, optomechanics and thermal camouflage [20-26] as well as metasail design [35-36]. Furthermore, the excitation of plasmonic modes of the metallic nanoresonators, utilized in the metasurfaces unit cells, creates strong near-field intensity, which in turn improves the light-graphene interaction. Additionally, the work presented here will be the foundation of tunable metasurfaces for nuclear medical imaging applications such as Positron Emission Tomography (PET) and Single Photon Emission Computed Tomography (SPECT) where scintillation light in the scintillator detectors needs to be guided towards photodetectors. The conventional mechanically pixelated scintillator arrays are being replaced with monolithic or semi-monolithic scintillators to achieve depth of interaction (DOI) information and improved photon collection efficiency. This has been an active area of research for PET and SPECT detectors where DOI information and collection efficiency are shown to improve using light manipulation techniques such as laser induced optical barriers, retroreflectors and photonic structured scintillators [28-34].

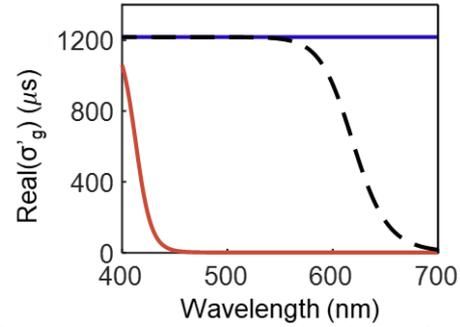

**Figure 1.** The real part of the total conductivity function for multi-sheet graphene layer $real(\sigma'_g)$ with $N = 20$. In this figure the impact of changing $\mu_c$ on the $real(\sigma'_g)$ is presented. The $real(\sigma'_g)$ is presented for graphene layers with Fermi levels of $\mu_c = 1.5 eV$, $\mu_c = 1 eV$ and $\mu_c = 0.25 eV$ which are exhibited by red, dashed black, and blue curves, respectively.

In this paper we theoretically studied the optical forces acting on a plasmonic metasurface integrated with a layer of graphene sheets, in the visible band, using the Finite-Difference Time-Domain (FDTD) method. The normal and tangential forces are discussed and computed via numerical simulation and analytical formulation of the Maxwell Stress Tensor (MST). By tailoring the properties of the metasurface unit cell, we can increase the anomalous reflection efficiency in the solar emission spectrum, where this property can be tuned using an active graphene layer to control and achieve the desired radiation forces, in the preferred direction.

## 2. Simulation setup and analytical formalism

Active metasurface configuration is a combination of (i) graphene layer (ii) plasmonic phase gradient unit cell and (iii) asymmetric Fabry-Perot cavity. In this section the optical properties of graphene are discussed and the reason for choosing a multi-sheet graphene layer is disclosed. Additionally, the design parameters of the Fabry-Perot cavity and plasmonic phase gradient unit cell, which is made of silver resonators are explained. Finally, the theoretical formalism for MST which is utilized alongside the numerical method to study the exerted forces is discussed.

### 2.1 Optical properties of graphene

Graphene provides a tunable optical response from terahertz to the visible band, which can be integrated with photonics. In our study graphene is modeled as a conductive sheet with the surface conductivity given by the Kubo formula as [17]

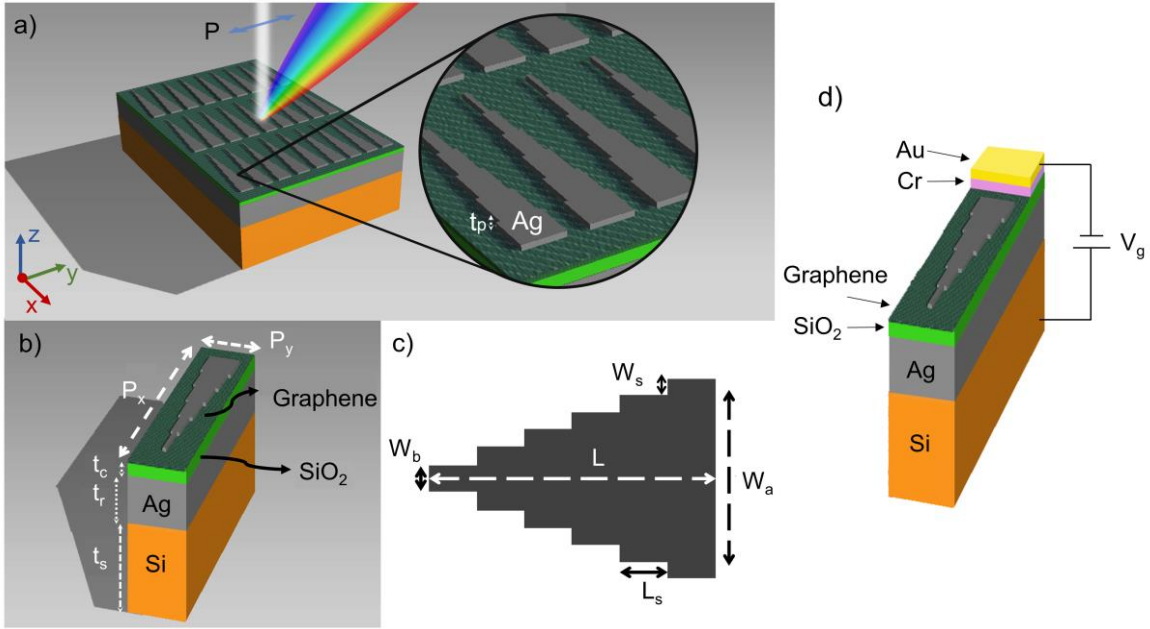

**Figure 2.** 3D schematic of the metasurface with an array of tapered silver unit cells on top of the graphene layer. (b) The unit cell stack of the metasurface depicting the structural dimensions. (c) Top view of the phase gradient structure with six silver nanoresonators positioned nearby without gap. (d) Gating approach with Cr and Au to control the Fermi level of the graphene sheets.

$$\sigma_g(\omega, \mu_c, \Gamma, T) = \sigma_{\text{intra}} + \sigma_{\text{inter}} \quad (1)$$

The first and second terms of eq. (1) are representing the intraband and interband contributions to the conductivity. Intraband contributions can be defined as:

$$\sigma_{\text{intra}} = -j\frac{e^2 k_B T}{\pi\hbar^2(\omega - j2\Gamma)}\left[\frac{\mu_c}{k_B T} + 2\ln\left(e^{-\mu_c/k_B T} + 1\right)\right] \quad (2)$$

Additionally, for $k_B T \ll |\mu_c|$ and $k_B T \ll \hbar\omega$, the interband conductivity is given by:

$$\sigma_{\text{inter}} = \frac{-je^2}{4\pi\hbar^2}\ln\left[\frac{2|\mu_c| - \hbar(\omega - j2\Gamma)}{2|\mu_c| + \hbar(\omega - j2\Gamma)}\right] \quad (3)$$

where e, $k_B$, $\mu_c$, and $T$ are electron charge, Boltzmann constant, Fermi level, and temperature in Kelvin, respectively. Furthermore, $\Gamma$ and $\hbar$ are the phenomenological scattering rate and reduced Planck constant, respectively. In the case of having $N$ graphene sheets, the total conductivity function of the graphene layer can be written as

$$\sigma'_g = N\sigma_g \quad (4)$$

where $\sigma_g$ is the conductivity function of the single graphene sheet, given by eq. (1). Previously the possibility of using multiple graphene sheets in a metasurface structure is discussed in [16], which reveals that it can enhance the tunability performance and anomalous reflection efficiency of the metasurface.

**Table 1.** Geometric parameters of the metasurface unit cell

| Parameter | $t_p$ | $t_c$ | $t_r$ | $t_s$ | $W_a$ | $W_s$ | $W_b$ | L | $L_s$ |
|---|---|---|---|---|---|---|---|---|---|
| Value (nm) | 30 | 80 | 170 | 750 | 170 | 15 | 20 | 990 | 165 |

In the visible light regime, the interband term of the conductivity function (i.e., $\sigma_{\text{inter}}$) has a significant contribution to the total conductivity of graphene. Moreover, the real part of the total conductivity function (i.e., $real(\sigma_g)$) in the mentioned spectra, which is mainly the contributions of interband absorption of the graphene band structure, spectrally blueshifts via increasing the electrochemical potential level ($\mu_c$). As demonstrated in figure 1, $N = 20$ (i.e., 20 graphene sheets in a graphene layer) and $\mu_c = 1.5eV$ lead to $real(\sigma'_g) \sim 1.2\ \mu S$, however, by having a lower Fermi level as $\mu_c = 0.25eV$ the real part becomes as large as $1200\ \mu S$, which is about 3 orders of magnitude larger at $\lambda_0 = 550nm$, and covers the entire visible light spectrum. This outstanding feature could be used in designing tunable metasurface reflectors and camouflage devices [26].

## 2.2 Metasurface design

The 3D schematic of the plasmonic metasurface is presented in figures 2(a) and 2(b). Each phase gradient unit cell is a quasi-continuous tapered silver cell, which consists of six rectangular nanoresonators positioned without gap, parallel to each other, along the y-axis, and with the gradual change in their width on the x-axis. The periodicity of the unit cell along the x and y-axis are $P_x = 1100\ nm$ and $P_y = 250\ nm$, respectively. The thickness of the silver phase gradient on the z-axis is $t_p = 30 nm$. A top view of the phase gradient unit cell is depicted in figure 2(c). This configuration provides a sensitive plasmonic phase gradient unit cell in the x-direction for the transverse electric (TE) polarized light in the y-direction. Furthermore, a graphene layer with 20 sheets is embedded under the nanoresonators and on top of the $SiO_2$ cavity. A fully reflective silver mirror is positioned under $SiO_2$ layer, creating an asymmetric Fabry-Perot cavity configuration, where the mirror is laying on top of the Si substrate. Dimensions of the metasurface unit cell, which were found to be optimum and are used in our study, are displayed in Table 1. We aim to combine continuous phase gradient configuration (which have trapezoid shapes) with discrete phase change unit cells (rectangular resonators) to enhance anomalous reflection efficiency and mitigate specular reflection.

In section 2.1, the optical properties of graphene are discussed extensively, and the conductivity of the graphene layer is given in eq. (4). Additionally, the refractive index of the $SiO_2$ is set to 1.45. For the frequency-dependent refractive index function of the silver, the Johnson and Christy tabulations are used [27].

The phase shift over the phase gradient unit cell for this structure along the x-axis is calculated via reflected light at the free space wavelength of $\lambda_0 = 550 nm$ and the results are depicted in figure 3. Since the thicknesses of the resonator and cavity (i.e., $t_p$ and $t_c$) are in deep subwavelength scales, the phase accumulation in the metasurface media is mainly due to the phase gradient structure, with the end-mirror adding a $\pi$ phase to the reflected light.

Based on generalized Snell's law we can obtain the phase gradient's operating bandwidth [17]. On the other hand, several factors are effective in determining the final operating bandwidth of the metasurface, such as material properties, or the shape of the metasurface antennas. Generalized Snell's law is given by [1],

$$sin(\theta_r) - sin(\theta_i) = \frac{\nabla \phi_x}{k_0} \quad (5)$$

where $k_0$ is the free space wavevector. As well, $\theta_r$ with $\theta_i$ are reflection and incident angles, respectively. In our theoretical analysis, we assume that the incident angle is zero (i.e., the incident light is normal to the plane of the metasurface, $\theta_i = 0°$). Also, $\nabla \phi_x = 2\pi/P_x$, where the $P_x$ refers to the periodicity of the metasurface unit cells in the phase gradient direction.

Here, according to the Generalized Snell's law, the operation bandwidth of the metasurface can reach up to $\lambda_0 = 1100 nm$. The shorter cutoff wavelength, on the other hand, is determined by the material properties of the silver resonators, which have the plasmonic resonance starting from $400 nm$ and it is dependent on the structural dimensions and shape of the silver resonators. Furthermore, placing the graphene layer underneath the silver resonators and tunability of the surface conductivity of graphene can affect the anomalous reflection efficiency of the metasurface ($\eta_e$).

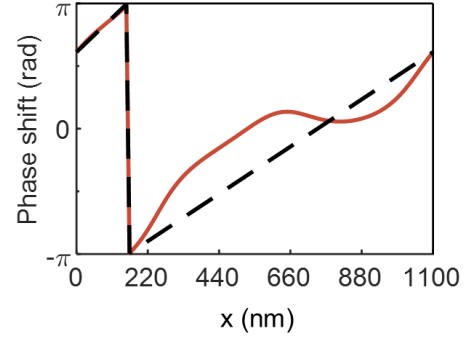

**Figure 3.** Phase shift over the unit cell of the metasurface at the free space wavelength of $\lambda_0 = 550 nm$. The red curve presents the metasurface in which the graphene's Fermi level is set to $\mu = 1.5 eV$. The linear phase shift fitting the unit cell phase change is presented with the black dashed curve.

Figure 4 presents how changing the Fermi level of the graphene layer ($\mu_c$) alters the metasurface anomalous reflection efficiency. The efficiency of the metasurface can be described by dividing anomalously reflected light by the total incident power. Here, the absorption of light by metasurface assembly, as well as the specular and parasitic reflections are considered as the deficiency of the metasurface. The normalized intensity of reflected light at $\lambda_0$=550 nm is represented in figure 4(a), for (i) without graphene layer (dashed black curve) (ii) with graphene layer of $\mu_c = 1.5 eV$ (red curve referred to as ON state) (iii) with graphene layer of $\mu_c = 0.25 eV$ (blue curve referred as OFF state). As the simulation results unveil, at the ON state, a considerable portion of the light is reflected anomalously, while the intensity of the specular and parasitic modes are almost ~10 times smaller, comparatively. The far-field projection intensity varies by placing the graphene layer (i.e., 20 sheets of graphene) underneath the phase gradient surface and on top of the asymmetric cavity. The conductivity of graphene is controlled by the Fermi level, and its adjusting changes the anomalous reflection efficiency.

Since the incident light is normal to the metasurface (i.e., $\theta_i = 0°$), at the incident light's wavelength $\lambda_0 = 550 nm$, the anomalous reflection angle can be calculated via (5), as $sin(\theta_r) = \nabla \phi_x/k_0 = 0.5$, (i.e., $\theta_r = 30°$). The anomalous reflection intensity for the metasurface with graphene layer and Fermi level of $\mu_c = 1.5 eV$ (i.e., red curve, referred as ON state) is even higher than the metasurface without graphene, (i.e., blue curve). Another important characteristic of the metasurface is revealed in figure 4 and when the Fermi level is $\mu_c = 0.25 eV$. Here not only the anomalous reflection intensity becomes low, but also absorption of light is more effective. Since by varying $\mu_c$ the interband absorption of the graphene layer spectrally shifts, and the general absorption coefficient $\alpha = 2.303A/2d$ of the metasurface is modified as well. Here, $d$ is the thickness of the cavity and phase gradient together. Absorption coefficients of the metasurface at $\lambda_0 = 550nm$ for $\mu_{c1} = 0.25eV$, and $\mu_{c2} = 1.5eV$ are calculated as $\alpha_1 = 8.1 \times 10^6\ m^{-1}$ (i.e., $A_1 = 78\%$ absorption) and $\alpha_2 = 1.3 \times 10^6 m^{-1}$ (i.e., $A_2 = 13\%$ absorption), respectively. In other words, by increasing $\mu_c$ the loss in the system for the visible spectra reduces, while reducing $\mu_c$ has the inverse influence (i.e., increasing the absorption of light).

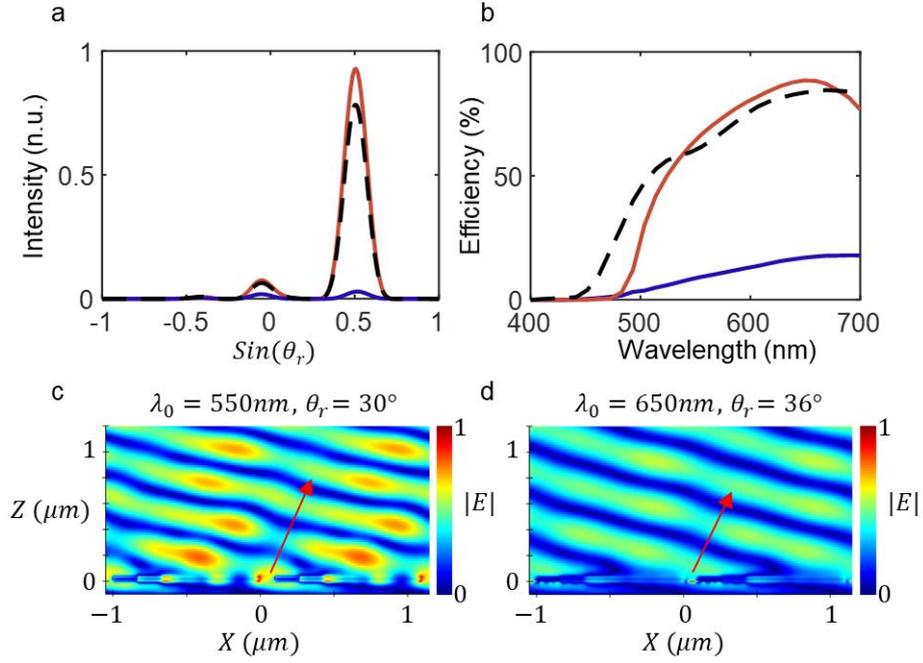

**Figure 4.** (a) Normalized reflection intensity for the incident light ($\theta_i = 0°$) at $\lambda_0 = 550nm$. (b) Anomalously reflected light efficiency ($\eta_e$) over the visible spectrum. The red curve presents the Fermi level of $\mu_c = 1.5eV$ (i.e., ON state), while the blue curve displays the metasurface with graphene layer of $\mu_c = 0.25eV$ (i.e., OFF state). The black dashed curve presents the metasurface without graphene. (c) and (d) are presenting the reflected light with free space wavelengths of $\lambda_0 = 550nm$ and $\lambda_0 = 650nm$, respectively. At longer wavelengths the angle of reflection becomes larger, which is in resemblance to the Snell's law.

Furthermore, in figure 4(b), one can observe the efficiency of the metasurface over the operating visible bandwidth. As it is disclosed, similar trend of figure 4(a) is also appeared in figure 4(b). With setting $\mu_c = 1.5eV$ the anomalous reflection efficiency in overall spectra increases as high as $\eta_e = 88\%$ due to loss reduction, and it becomes $\eta_e = 66\%$ at $\lambda_0 =550nm$. As mentioned previously, decreasing the Fermi level to $\mu_c = 0.25eV$ enhances the absorption, decreasing the anomalous reflection efficiency to $\eta_e = 9\%$ at $\lambda_0 =550nm$. Hence, the proposed structure behaves as a tuneable reflector/absorber. Figure 4(c) and (d) are presenting the reflected field for various incident wavelengths. As the incident wavelength becomes longer, the angle of reflection obtains larger values, which is in accordance with Snell's law

(5). The metasurface is at ON state (i.e., $\mu_c = 1.5eV$) in figures 4(c) and 4(d).

## 2.3 Optical forces acting on metasurface

We extensively studied the optical forces acting on the metasurface, which are induced by the incident light normal to the metasurface plane. In the specular reflecting surfaces, since $\theta_i = \theta_r$, the tangential forces are in opposite directions, eliminating each other (i.e., $F_\parallel = 0$), while the normal component of the force propels the surface (i.e., $F_\perp \neq 0$). In the anomalously reflecting surfaces, given that $\theta_i \neq \theta_r$, the tangential forces of incoming and reflecting light do not cancel out each other, which results in the creation of the tangential force in these structures, additional to the normal component of the force (i.e., $F_\parallel \neq 0$ and $F_\perp \neq 0$). Here, MST is used to compute the optical forces acting on the metasurfaces [6], [7].

$$\langle F \rangle = \oint_v \langle \nabla \cdot \overleftrightarrow{T} \rangle \cdot dv = \int_s \langle \overleftrightarrow{T} \rangle \cdot \hat{n}\, ds \qquad (6)$$

where $\overleftrightarrow{T}$ is the MST and $\hat{n}$ is the normal vector to the integrating surface $s$. The time-averaged force is calculated by integrating the MST over the surrounding closed surface. The time-averaged MST is given by

$$\langle \overleftrightarrow{T} \rangle = D \otimes E^* + B \otimes H^* - \frac{1}{2} I (D.E^* + B.H^*) \quad (7)$$

where $E$ and $H$ are electric and magnetic fields, respectively. $D$ and $B$ denote the electric displacement and magnetic flux density, respectively, and $I$ is the identity matrix. The net force exerted on the metasurface is the sum of all acting forces due to the incident (i.e., a force due to the incident $\langle F \rangle_i$) and reflected light (i.e., a force due to the reflection $\langle F \rangle_r$), which are accountable for propulsive force $\langle F_\perp \rangle$ (i.e., $\langle F_z \rangle$), and tangential force $\langle F_\parallel \rangle$ (i.e., $\langle F_x \rangle$). As a result, the net forces acting on the metasurface are given by,

$$\langle F_\perp \rangle_{Net} = \langle F_\perp \rangle_i + \langle F_\perp \rangle_r \quad (8)$$

$$\langle F_\parallel \rangle_{Net} = \langle F_\parallel \rangle_i + \langle F_\parallel \rangle_r \quad (9)$$

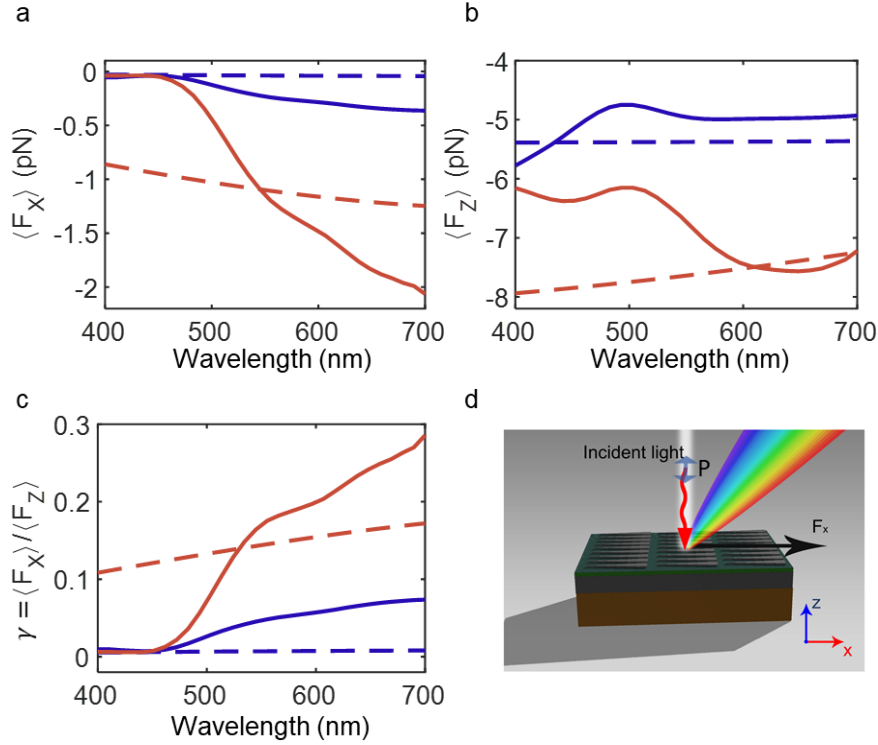

**Figure 5.** (a) Tangential forces acting on metasurface (i.e., $\langle F_x \rangle$). red and blue solid lines are obtained from simulation and calculation of MST, which are presenting ON ($\mu_c = 1.5eV$) and OFF ($\mu_c = 0.25eV$) states, respectively. The dashed red and blue lines are computed forces with analytical formalism, which present ON and OFF states, respectively. (b) propulsive forces acting on metasurface (i.e., $\langle F_z \rangle$). Like (a) red and blue curves are presenting the ON and OFF states, while solid and dashed curves are for simulation and theoretical modelling, respectively. (c) The ratio of the tangential force to normal force acting on the metasurface ($\gamma$), for ON and OFF states.

Here for the simplicity, we only study the radiation pressure induced by the normal incident (i.e., $\theta_i = 0, \langle F_\parallel \rangle_i = 0$). Force due to reflected light, can have components with different directions: (i) $\langle F_\perp \rangle_{r,spec}$ is a normal force induced by specular reflection, while (ii) $\langle F_\parallel \rangle_{r,anom}$ the tangential force induced by the anomalous reflection, and (iii) $\langle F_\perp \rangle_{r,anom}$ the normal force induced by anomalous reflection. It should be mentioned that the anomalously reflected light exerts force in both tangential and normal directions, with the magnitude of each being dependent on the reflection angle and in other words to $\zeta$.

The normal and the tangential components of the optical force induced by a TE polarized (S-polarized) plane wave incident light can be calculated from MST [7] and given by,

$$\langle F_\perp \rangle_i = -\frac{1}{2} \varepsilon_0 L_i L_j E_0^2 \cos^2(\theta) \hat{z} \quad (10)$$

$$\langle F_\parallel \rangle_i = \frac{1}{4} \varepsilon_0 L_i L_j E_0^2 \sin(2\theta) \hat{x} \quad (11)$$

where, $\theta$ refers to the incident angle. Also, the normal and the tangential components of the reflected light are:

$$\langle F_\perp \rangle_r = -\frac{L_i L_j E_0^2}{4\eta_0^2 k_0^2} \left( \mu_0(-2\zeta^2 - 4\zeta k_0 \sin(\theta) + k_0^2 \cos(2\theta)) + \eta_0^2 k_0^2 \varepsilon_0 \right) \hat{z} \quad (12)$$

$$\langle F_\parallel \rangle_r = -\frac{\mu_0 L_i L_j E_0^2}{2\eta_0^2 k_0} (\zeta + k_0 \sin(\theta)) \sqrt{1 - \left(\sin(\theta) + \frac{\zeta}{k_0}\right)^2} \hat{x} \quad (13)$$

Here $L_i$ and $L_j$ are the lengths of the square area on the metasurface receiving optical pressure, and $\theta$ refers to the incident light angle. Additionally, $\eta_0$, $\mu_0$ and $\varepsilon_0$ are free space wave impedance, free space permeability, and free space permittivity, respectively. It must be noted that $\langle F_z \rangle = \langle F_\perp \rangle_{Net}$ and $\langle F_x \rangle = \langle F_\parallel \rangle_{Net}$, which are alternative expressions. Considering the impact of the anomalous reflection efficiency ($\eta_e$) on the reflected light's induced force, and also considering the influence of absorption ($A$), we can rewrite normal and tangential forces for $\theta = 0°$. The value $1 - A$ becomes a significant factor for determining the net forces where by considering $A = 1$, $\langle F_z \rangle = \langle F_\perp \rangle_i$ and $\langle F_x \rangle = 0$. On the other hand, $\eta_e$ impacts both $\langle F_z \rangle$, $\langle F_x \rangle$ where we have $\zeta$ term, which is determining the ratio of reflected light's induced force acting in x-direction or z-direction. In general, we can write $\langle F_x \rangle = \langle F_\parallel \rangle_i + (1-A)\left((1-\eta_e)\langle F_\parallel \rangle_{r,spec} + (\eta_e)\langle F_\parallel \rangle_{r,anom}\right)$, similar case is for $\langle F_z \rangle$. If any of $\eta_e = 0$ or $\zeta = 0$ the $\langle F_z \rangle = \langle F_\perp \rangle_i + \langle F_\perp \rangle_{r,spec}$ and $\langle F_x \rangle = 0$, which simultaneously means that there is no anomalous reflection.

Thus, the net forces in $\hat{z}$ and $\hat{x}$ directions by considering $\theta = 0°$ (i.e., $\langle F_\parallel \rangle_i = 0$) are given by:

$$\langle F_z \rangle = \left(-\frac{1}{2}\varepsilon_0 L_i L_j E_0^2 + (1-A)\frac{L_i L_j E_0^2}{4\eta_0^2 k_0^2}\left(\mu_0(-2\eta_e \zeta^2 + k_0^2) + \eta_0^2 k_0^2 \varepsilon_0\right)\right)\hat{z} \quad (14)$$

$$\langle F_x \rangle = -(1-A)\frac{\mu_0 L_i L_j E_0^2 \eta_e \zeta}{2\eta_0^2 k_0}\sqrt{1-\left(\frac{\zeta}{k_0}\right)^2}\hat{x} \quad (15)$$

## 3. Main results and discussion

Figures 5(a) and 5(b) compare the total tangential and normal forces acting on the metasurface with different Fermi levels of graphene. The red curve presents optical force for the case with $\mu_c = 1.5 eV$ (The ON state), while the blue curve presents the optical force for the case with $\mu_c = 0.25 eV$ (The OFF state). Since at the OFF state the metasurface does not present significant anomalous reflection (i.e., low $\eta_e$) and has higher absorption, we obtain smaller tangential force (i.e., smaller $\langle F_x \rangle$) and smaller normal force (i.e., smaller $\langle F_z \rangle$). On the other hand, at the ON state, since the anomalous reflection is stronger (i.e., higher $\eta_e$) and absorption is less, the obtained tangential force and normal force are larger, respectively. Figure 5(d) presents the ratio of the tangential force to the normal force $\gamma = \langle F_x \rangle / \langle F_z \rangle$. The larger value of $\gamma$ at the ON state reveals that the metasurface produces greater tangential force in the total momentum harvesting process. This mechanism in turn can provide tunability in the displacement and acceleration direction.

Furthermore, the theoretical forces are presented with the red dashed curve, and blue dashed curve for ON and OFF states, respectively. In theoretical formalism constant values for $\eta_e$ and $A$ for full spectra of ON and OFF states are considered to avoid complexity. We aimed $\eta_e = 66\%$ and $A = 13\%$ for the ON state and $\eta_e = 9\%$ and $A = 78\%$ for the OFF state, where the $\eta_e$ and $A$ at $\lambda_0 = 550 nm$ are obtained from simulation data. Optical forces in both theoretical and simulation methods are calculated for an area equal to $1.1 \times 10^{-12}\ m^2$, which is 4 times larger than the area of the unit cell (i.e., $0.275 \times 10^{-12}\ m^2$), however, for simplicity, optical forces are normalized for an area as big as $1\ m^2$.

With a longer incident wavelength, according to Snell's law, the reflected light's angle becomes larger, therefore $\langle F_x \rangle$ and $\langle F_z \rangle$ become larger and smaller, respectively. For the same reason at longer wavelengths, $\gamma$ obtains larger values. Furthermore, the slight mismatch between trends of the theoretical and numerical results can be clarified as follows: in the numerical method the $\eta_e$ and $A$ values are in continuous variation over the spectrum, while in the numerical model we used constant values. Despite these variations, the overall magnitude of the numerical and theoretical formalisms follows each other exhibiting to the credibility of the methods.

## 4. Conclusion

In this study, we presented a metasurface with tunable anomalous reflection/absorption, providing controllable optical forces acting on the metasurface. The anomalous reflection efficiency of the metasurface at $\lambda_0 = 550 nm$ reaches $\eta_e = 78\%$ when the metasurface is at the ON state (i.e., when the graphene's Fermi level is $1.5 eV$) and it is $\eta_e = 13\%$ for the OFF state (i.e., when graphene's Fermi level is $0.25 eV$), which in turn provides the tunable tangential and normal forces acting on the surface. It is disclosed that by changing the system from the OFF state to the ON state, the ratio of the tangential forces to the normal forces becomes 4 times bigger. Feasibility in tuning the magnitude of the force in various directions, and the possibility to integrate metasurface with multiple phase gradients units in distinct directions, can aid us in deterministically exerting forces in diverse chosen directions, and therefore move objects in the desired route or relocate them spatially. Additionally, the idea of tunable metasurfaces can be used in healthcare and biological systems as well as in designing new generation of tunable SPECT and PET scintillators.

## Acknowledgements

S.S.M., A.A., and Si.S. conceived the idea and performed simulations. Se.S., F.B., and H.S. provided theoretical contributions to this research. Si.S performed theoretical analysis and modelling to support the idea. All authors contributed to writing and organizing the manuscript.

## Funding

This work was not supported by any funding sources.

## Data availability statement

All data supporting the findings of this research are included within the article.